\documentclass[amsmath,amssymb,amsbsy,prl,twocolumn,superscriptaddress,showpacs]{revtex4}
\usepackage{amsfonts}
\usepackage{amsmath}
\usepackage{amssymb}
\usepackage{bm}
\usepackage[usenames, dvipsnames]{color}
\usepackage{dcolumn}
\usepackage{epsfig}
\usepackage{eucal}
\usepackage{graphicx}
\usepackage[breaklinks,colorlinks = true,linkcolor = red,urlcolor=cyan,citecolor=red]{hyperref}
\usepackage{mathrsfs}
\usepackage{subfigure}
\usepackage{textcomp}
\usepackage{times}
\usepackage{xy}

\newcommand{\ourtitle}{Orbital Selective Superconductivity in a Two-band Model of Infinite-Layer Nickelates}

\begin{document}

\title{\ourtitle}
\author{Priyo Adhikary$^{\ast}$}
\affiliation{Department of Physics, Indian Institute of Science, Bangalore 560012, India.}
\author{Subhadeep Bandyopadhyay$^{\ast}$}
\affiliation{School of Physical Sciences, Indian Association for the Cultivation of Science, Kolkata 700 032, India.}
\author{Tanmoy Das}
\email{tnmydas@gmail.com}
\affiliation{Department of Physics, Indian Institute of Science, Bangalore 560012, India.}
\author{Indra Dasgupta}
\email{sspid@iacs.res.in}
\affiliation{School of Physical Sciences, Indian Association for the Cultivation of Science, Kolkata 700 032, India.}
\author{Tanusri Saha-Dasgupta}
\email{t.sahadasgupta@gmail.com}
\affiliation{S. N. Bose National Centre for Basic Sciences, JD Block, Sector III, Salt Lake, Kolkata, West Bengal 700106, India.}

\begin{abstract}
  In the present study, we explore superconductivity in NdNiO$_2$ and LaNiO$_2$ employing a first-principles derived low-energy model Hamiltonian,
  consisting of two orbitals: Ni $x^{2}$-$y^{2}$, and an {\it axial} orbital. The {\it axial} orbital is constructed out of Nd/La $d$,
  Ni 3$z^{2}$-$r^{2}$ and Ni $s$ characters. Calculation of the superconducting pairing symmetry and pairing eigenvalue of the
  spin-fluctuation mediated pairing interaction underlines the crucial role of inter-orbital Hubbard interaction in superconductivity,
  which turns out to be orbital-selective. The axial orbital brings in materials dependence in the problem, making NdNiO$_2$ different
  from LaNiO$_2$, thereby controlling the inter-orbital Hubbard interaction assisted superconductivity.
\end{abstract}  
\maketitle

{\it Introduction.--} The discovery of high T$_c$ superconductivity in cuprate family,\cite{cuprate} has kept the researchers busy over the last two decades or so, in understanding the mechanism of its superconductivity. While the mechanism of superconductivity in cuprates still remains debated, attempt has been made in search of superconductivity in
transition-metal oxides other than cuprates. Nickelates, Ni being neighboring element to Cu, turns out to be most promising in this respect. Towards this goal, a number of attempts has been made. For example, LaAlO$_3$/LaNiO$_3$ superlattice\cite{lao-lno} with theoretically predicted singly occupied Ni $x^{2}$-$y^{2}$ band,\cite{oka,millis} bilayer and
trilayer reduced La$_3$Ni$_2$O$_6$ and La$_4$Ni$_3$O$_8$\cite{martha, tsd, pickett-multi} have been studied. Unfortunately, experimental realization of superconductivity in these systems remained elusive. In this
connection, stabilization of unusual Ni$^{1+}$ valence in LaNiO$_2$ and NdNiO$_2$, having the same 3d$^9$ configuration as Cu$^{2+}$ in cuprates, in the isostructural infinite-layer CaCuO$_2$ structure,\cite{lno,nno} therefore generated interest. Recent report\cite{hwang} of superconductivity in 20$\%$ Sr doped NdNiO$_2$ with T$_c$ of
9-15 K has reignited this interest with reports of several theoretical studies \cite{aoki,rony,millis-nno,karsten,JiangDFT,GuDFT,AritaDFT,pickett,Devereaux,Changdwave,ZhangKondo,Vishwanath,SavrasovDMFT} devoted to understand this.

The most sought after issue, in this respect, has been whether Nd/LaNiO$_2$ (N/LNO) can be described within the same theoretical framework as cuprates. It is noteworthy that while the undoped cuprates are antiferromagnetic with strong insulating properties,\cite{cuprate-afm} undoped N/LNO is reported to be nonmagnetic with bad metallic properties.\cite{lno,hwang} Comparing the electronic structure with that of cuprates, the most obvious difference is the large O 2$p$ - Ni 3$d$ charge transfer energy for Ni$^{1+}$ compared to small O 2$p$- Cu 3$d$ charge transfer energy for Cu$^{2+}$, resulting in a significantly weaker hybridization with O 2$p$ compared to cuprates. This fact has been evidenced in electronic structure calculations\cite{millis-nno,pickett} as well as in terms of the absence of pre-peak in x-ray absorption spectroscopy (XAS) near the O K-edge.\cite{sawatzky} This also provides justification for weak super-exchange in NNO compared to strong antiferromagnetism in undoped cuprates, the super exchange energy scale being estimated to be a factor of 10 smaller compared to cuprates.\cite{J-scale} The large charge transfer energy puts the Ni 3$d$ levels higher up in energy compared to Cu 3$d$, which facilitates their hybridization with usually empty Nd/La $d$ bands. While the hybridization of in-plane Ni $x^{2}$-$y^{2}$ with out-of-plane Nd/La $d$ orbitals is negligible, they can hybridize via the Ni $3z^{2}$-$r^{2}$ and Ni $s$.\cite{pickett, SIT} This creates two three-dimensional (3D) Fermi surface (FS) pockets in N/LNO, one centered around $\Gamma$ and another around A point of the 3D BZ.\cite{rony,pickett} The Nd/La-Ni hybridization has been signaled in additional low energy shoulder features in XAS and resonant inelastic x-ray scattering (RIXS) spectra of N/LNO.\cite{sawatzky} Fitting of the XAS and RIXS spectra, through exact diagonalization calculation produced a mixing from rare-earths as large as 44 $\%$.\cite{sawatzky} Thus while in case of cuprates, the intercalated non-copper-oxide layers act as spectators, playing the role of simple “charge reservoir”, in case of N/LNO the rare-earth layer may provide active electronic degrees of freedom. This prompted Hepting et al.\cite{sawatzky} to suggest a two-band model, consisting of a 3D rare-earth band coupled to a Ni $x^{2}$-y$^{2}$ derived 2D Mott system. Interestingly, a more recent report shows a sign reversal of Hall resistivity with doping and temperature,\cite{hwang-new} which may further evince the two band scenario. 

Given this background, a natural question would be, can doped carriers give rise to superconductivity in such a two-band model description of NNO and LNO, and if so, is there any difference between NNO and LNO. This, to the best of our knowledge, has remained unexplored so far, though superconductivity in nickelates has been explored within the framework of one-band, and three-band model with onsite correlations,\cite{rony} one band Hubbard model,\cite{karsten} and multi-band Ni $d$ - Nd $d$ Hubbard model within fluctuation-exchange approximation,\cite{aoki} as well as from strong coupling starting points.\cite{Vishwanath,Changdwave,ZhangKondo} 

In the present study, we first construct a two-band model, starting from the self-consistent-field density functional theory (DFT) and by retaining Ni $x^{2}$-$y^{2}$ and Ni $s$ orbitals in the basis, and downfolding the rest. Here, our main finding is that the Ni $s$ basis forms an axial orbital, resulting from the hybridization of Nd/La 3$z^{2}$-$r^{2}$, Nd/La $xy$, Ni 3$z^{2}$-$r^{2}$ and Ni $s$. Moreover, while the downfolded Ni-$x^{2}$-$y^{2}$ Wannier orbital is very similar in the La and Nd compounds, the detailed nature of the axial orbital set these two materials apart, giving clue to its possible role on the materials dependent superconductivity.

We next solve the pairing eigenvalue and pairing eigenfunctions of the spin-fluctuation mediated pairing interaction, computed within the DFT-derived two Wannier orbital Hubbard model. We find that (i) in the Nd compound, the superconducting (SC) coupling constant $\lambda$ grows almost exponentially with the inter-orbital interaction $V_{sd}$, while the intra-orbital interactions alone is not conducive for superconductivity. In a crude analogy with the renormalization theory, we can say that intra-orbital interactions are `marginal' $-$ do not directly mediate superconductivity, while the inter-orbital  interaction is a `relevant' parameter for superconductivity.  (ii) Secondly, in NNO, we find that the pairing eigenfunction turns out to be orbital selective: being a 2D $x^2$-$y^2$-type for the Ni $d$ orbital, and a 3D 3${z^2}$-$r^{2}$-type symmetry for the axial orbital. The results are consistent with the corresponding orbital weight distributions on the 3D FS topology, and the corresponding FS features. The same study in LNO results in a single $x^2$-$y^2$ wave pairing symmetry, but with SC coupling constant significantly smaller than that of NNO. Our findings emphasize the importance of axial orbital and a two-band model in which orbital selective pairing symmetry is augmented by the inter-orbital interaction.

\begin{figure}
\begin{center}
\rotatebox{0}{\includegraphics[width=0.5\textwidth]{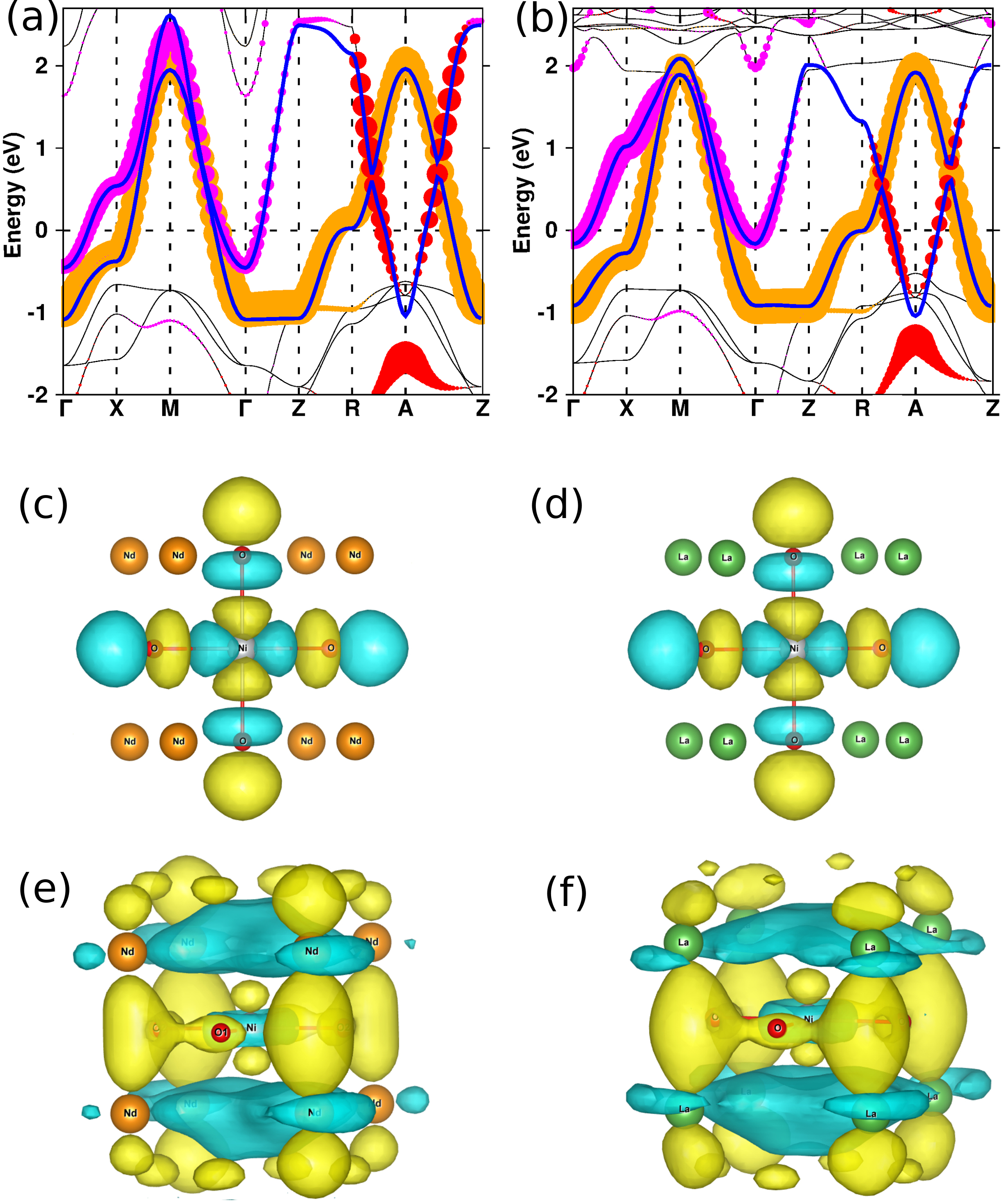}}
\end{center}
\caption{(Color online) (a)-(b) The DFT band structure (thin, black) together with downfolded two-band structure (thick, blue) for NNO and LNO, plotted along the high symmetry points of tetragonal unit cell, $\Gamma$(0,0,0)-X($\pi/a$,0,0)-M($\pi/a$,$\pi/a$,0)-
  $\Gamma$-Z(0,0,$\pi/c$)-R($\pi/a$,0,$\pi/c$)-A($\pi/a$,$\pi/a$,$\pi/c$)-Z. The fatness in DFT band structure corresponds to Ni $x^{2}$-$y^{2}$ (orange), Nd 3$z^{2}$-$r^{2}$ (magenta) and Nd $xy$ (red). (c)-(d) The Ni $x^{2}$-$y^{2}$ Wannier function in the downfolded two-band basis for NNO and LNO. (e)-(f) The Wannier function for axial orbital in the downfolded two-band basis for NNO and LNO. Plotted are the constant value surface with lobes of different signs colored as yellow and cyan.}
\label{fig1}
\end{figure}

{\it DFT band-structure and two-band model.--} The DFT band structure is computed in plane wave basis, as implemented in Vienna Ab-initio Simulation Package (VASP)\cite{vasp} with projected augmented wave (PAW) potential\cite{paw} and choice of generalized gradient approximation\cite{pbe} (GGA) for exchange-correlation functional. For details of the calculation, see supplementary materials (SM).\cite{suppl}
The resuts for undoped N/LNO is shown in Fig. 1(a)-(b). The band structure of N/LNO, which is well studied in literature,\cite{rony,pickett,arita,botana} primarily consists of O-2$p$ dominated bands ranging from about -8 eV to about -3 eV, Ni-3$d$ dominated bands ranging from about -3 eV to 2 eV, and Nd/La-5$d$ dominated bands ranging from about -1 eV to 8 eV. The low-energy electronic structure has two bands crossing the Fermi level: one canonical Ni $x^{2}$-$y^{2}$ band creating a hole pocket centered around M (A) point, bearing strong resemblance with cuprates, and the other one is derived out of Nd/La $d$ mixed with Ni characters creating electron pockets at $\Gamma$ and A points. While the generic features are found to be similar in the band structures of NNO [(a)] and LNO [(b)], there are subtle differences. Comparing the Ni $x^{2}$-$y^{2}$ bands in the two compounds, while it extends from -1.1 eV to  2 eV for NNO, it extends from -0.9 eV to about 2 eV for LNO, making the band width of Ni $x^{2}$-$y^{2}$ in LNO band smaller by about 0.2 eV as compared to NNO. The corresponding $k_z$ dispersion is also smaller for LNO compared to NNO. The saddle point at R is positioned about 0.2 eV higher compared to that at X in LNO, whereas R saddle point is about 0.5 eV higher compared to that at X for NNO. This $k_z$ dispersion highlights the mixing with the axial orbital, making the Ni $x^{2}$-$y^{2}$ band deviating from its 2D nature, as emphasized by Lee and Pickett.\cite{pickett} Comparing the second band, we find that firstly the Nd $d$-Ni derived electron pocket centered around $\Gamma$ is about -0.4 eV lower in energy in NNO as compared to LNO, making the self-doping effect more pronounced in the Nd compound compared to the La compound. Secondly, the width of the second band is about 1 eV smaller in LNO compared to NNO.

These subtle but important material-specific differences of the electronic structure of nickelate compounds get reflected in Wannier functions\cite{WAN} defining the downfolded two-band structure, designed to reproduce the two low energy bands of the DFT band structure. In order to construct the low energy two-band structure, we retain Ni $x^{2}$-$y^{2}$ and Ni $s$ degrees of freedom, and downfold the rest. This choice is guided by the four band model of cuprates,\cite{our,our1} consisting of Cu $s$, Cu $x^{2}$-$y^{2}$, O $p_x$ and O $p_y$, the material dependence being included in the Cu $s$ in downfolded basis, which forms the {\it axial orbital} by combining Cu $s$, Cu 3$z^{2}$-$r^{2}$, apical oxygen $p_z$
and orbitals of farther axial cations. In the present case, we downfold O $p_x$ and O $p_y$, as the larger charge transfer energy between Nd $d$ and O $p$ in nickelates, compared to cuprates, makes their mixing with $x^{2}$-$y^{2}$ much smaller than in cuprates. The resultant Wannier functions corresponding to downfolded two-band structure are shown in Fig. 1(c)-(f) for N/LNO. The $x^{2}$-$y^{2}$ Wannier function which forms $pd\sigma$ antibonding combination [(c)-(d)] is found to be identical between Nd and La compounds. We note that the $p$-like tail of $x^{2}$-$y^{2}$ Wannier functions sitting at O sites show asymmetry between the positive and negative lobes, which signifies the mixing with diffuse Ni $s$. The material dependence, however, shows up in the Wannier function which forms the axial orbital [(e)-(f)]. We find this axial orbital is a hybrid between Ni $s$, Ni 3$z^{2}$ - $r^{2}$, Nd/La 3$z^{2}$ - $r^{2}$ and Nd/La $xy$. Inspecting this orbital, we find that starting from the central Ni atom, Ni 3$z^{2}$ - $r^{2}$ which bonds to Ni $s$, and antibonds to O $p_x/p_y$, bonds strongly with predominant feature of
Nd/La 3$z^{2}$ - $r^{2}$ and $xy$, highlighting the hybridization between Ni and Nd/La $d$. We find that the Ni 3$z^{2}$ - $r^{2}$/Ni $s$ character is more in La compound [(f)] compared to Nd compound [(e)], while Nd 3$z^{2}$ - $r^{2}$/Nd $xy$ character is less in La compound compared to Nd compound. This makes the axial orbital much more cylindrical in NNO [(e)] and more spherical in LNO [(f)]. This differential nature of the axial orbital is reflected in both in-plane and out-of-plane hopping interactions within the two-band description (see SM) as well
as the energy of the axial orbital ($\epsilon_s$) measured from energy of $x^{2}$-$y^{2}$ ($\epsilon_d$) lying 0.25 eV higher in NNO compared to that in LNO. The in-plane hopping is found to be 30-20$\%$ larger in NNO compared to LNO. The out-plane hopping, especially the hopping connecting axial orbital to axial orbital show significantly larger values for NNO compared to LNO (1.2 to 7 times), accounting for about 1 eV larger band width of the axial band in NNO compared to LNO.

\begin{figure}[t]
\includegraphics[width=0.5\textwidth]{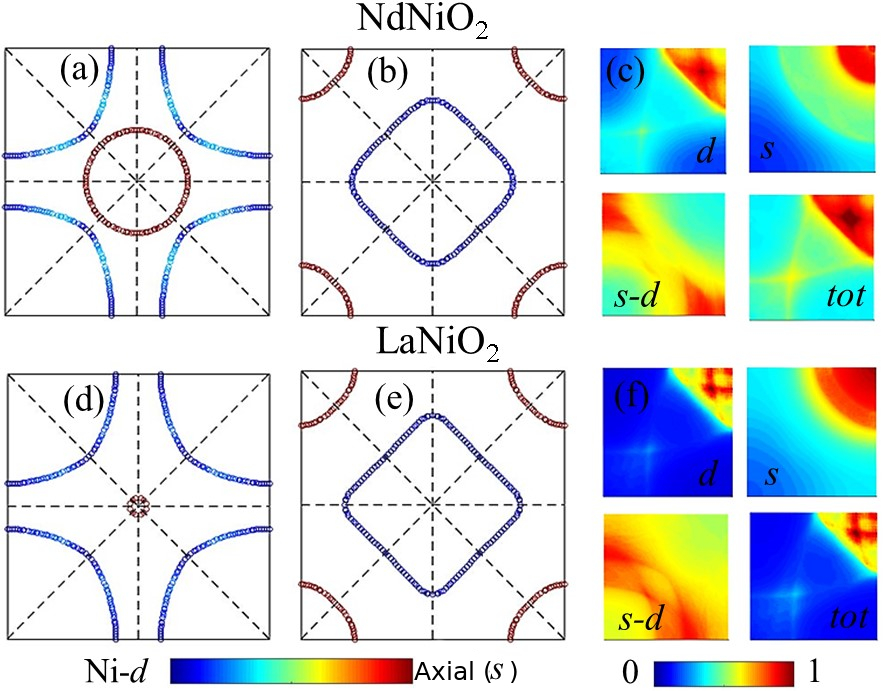}
\caption{(Color online) (a)-(b) FS topologies in NNO plotted as a function of $k_x-k_y$ [$-\pi\rightarrow\pi$ range] at
  $k_z$=0, and $\pi$. Blue (Ni $x^{2}$ - $y^{2}$) to red (axial) colors depict the orbital contributions at each $k_F$.
  (c) Plots of static spin susceptibility for two intra-orbital, inter-orbital, and total [Tr $\tilde{\chi}_s$] channels, for
  $q_z=\pi$, and $q_x,q_y$ :0 $\rightarrow$ $\pi$. All color bars are separately normalized for visualization. (d)-(f) Same as in (a), (b) and (c)
  but plotted for LNO.} 
\label{fig:FS} 
\end{figure}

{\it Calculation of superconducting properties.--} In analogy with cuprates,\cite{SCcuprates} pnictides,\cite{SCpnictides} and heavy-fermion superconductors,\cite{SCHF} we assume superconductivity in the present compound is spin-fluctuation mediated. The estimated electron-phonon interaction turns out to be too small to support observed T$_c$.\cite{arita} Based on a two-band Hubbard model, we obtain the pairing potential by considering the bubble and ladder diagrams:\cite{SCcuprates,SCpnictides,SCHF,SCrepulsive} 
\begin{eqnarray}
\tilde{\Gamma}({\bf q})&=&\frac{1}{2}\big[3{\tilde U}_{s}{\tilde \chi}_{s}({\bf q}){\tilde U}_{s} - {\tilde U}_{c}{\tilde \chi}_{c}({\bf q}){\tilde U}_{c} + {\tilde U}_{s}+{\tilde U}_{c}\big].
\label{singlet}
\end{eqnarray}
The symbol `tilde' denotes a tensor in the orbital basis. The subscripts `s' and `c' denote spin and charge fluctuation channels, respectively. $\tilde{U}_{s/c}$ are the onsite interaction tensors for spin and charge fluctuations, respectively whose non-vanishing components are $(\tilde{U}_{s,c})_{\alpha\alpha}^{\alpha\alpha}=U_{d/s}$ for intra-orbital $x^{2}$ - $y^{2}$ and axial, and the inter-orbital component is  $(\tilde{U}_{s,c})_{\alpha\alpha}^{\beta\beta}=V_{sd}$ ($\alpha\ne\beta$ are orbital indices).\cite{footnote1} $\tilde{\chi}_{s/c}$ are the spin and charge density-density correlation functions (tensors in the same orbital basis), computed within the random-phase-approximation (RPA). The details of the formalism is given in SM.\cite{suppl} Application of a weak coupling theory may be justified by the fact that exchange-scale in nickelates are smaller than cuprates.

We compute the eigenvalue and eigenfunctions of the pairing interaction $\tilde{\Gamma}({\bf k}-{\bf k}')$ on the 3D Fermi momenta, by solving the following equation:
\begin{eqnarray}
\Delta_{\nu}({\bf k})= -\lambda\frac{1}{\Omega_{\rm BZ}}\sum_{\nu',{\bf k}'}\Gamma'_{\nu\nu'}({\bf k}-{\bf k'})\Delta_{\nu'}({\bf k'}).
\label{SC2}
\end{eqnarray}
Here $\nu$, $\nu'$ denote band indices, and $\Gamma'_{\nu\nu'}$ is the pairing interaction, projected onto the band basis. $\lambda$ is the pairing eigenvalue (proportional to the SC coupling strength), and $\Delta_{\nu}({\bf k})$ is the corresponding pairing eigenfunction. Since the pairing potential is repulsive here, the highest {\it positive} eigenvalue $\lambda$, and the corresponding pairing symmetry can be shown to govern the lowest Free energy value in the SC state.\cite{SCrepulsive} 

The origin of the unconventional pairing symmetry, and the role of FS nesting can be understood as follows. For $\Gamma>0$ and $\lambda>0$ in Eq. 2, the pairing symmetry $\Delta_{\nu}({\bf k})$ must change {\it sign} over the FS to compensate for the negative sign in the left hand side of Eq. 2. $\Delta_{\nu}({\bf k})$ changes sign between the two ${\bf q}={\bf k}-{\bf k}'$-points, and either between different or same bands which are connected by the momentum ${\bf q}$ at which $\Gamma'_{\nu\nu'}({\bf q})$ acquires strong peaks. The locii of the peaks in $\Gamma'_{\nu\nu'}({\bf q})$ are primarily dictated by the FS nesting, while the overall amplitude is determined by $U_{s,d}$ and $V_{sd}$. We fix the hole doping level at $x=0.2$, which is about the optimal doping for NNO.\cite{hwang,hwang-new}

In Fig.~\ref{fig:FS}, we show the FS topology for NNO [(a)-(b)] and LNO [(d)-(e)] at two $k_z$ cuts, with the corresponding orbital weight indicated by red to blue color map. The FS-s are seen to be strongly 3D, which is typically detrimental for FS nesting strength. However, owing to the particular orbital weight distributions, there arise dominant nesting channels, which are highly orbital resolved. Interestingly, there is a complete orbital inversion among two FS sheets between $k_z$ = 0 and $\pi$. While the large hole pocket centering the zone boundary, and electron pocket in zone center of NNO BZ is of Ni $d$ ($x^{2}$-$y^{2}$) and axial character ($s$), respectively in $k_z$ = 0, they reverse their roles in $k_z$ = $\pi$. 
The Ni $d$ orbital enjoys a FS topology similar to the cuprates case in the low $k_z$ region, giving a nearly 2D FS nesting feature around ${\bf Q}=(\pi,\pi,0)$ and hence a $d_{x^2-y^2}$-pairing symmetry. On the other hand, the axial orbital acquires a FS nesting, considerably weaker in strength compared to the Ni $d$ orbital case, at  ${\bf Q}=(\pi,\pi,\pi)$, which is responsible for the 3${z^2}$-$r^{2}$ type pairing symmetry.
The FS for LNO, shown in Fig.~\ref{fig:FS}(d)-(e) is topologically similar to NNO, except it almost looses its FS pocket at the $\Gamma$-point. Since this heavily weakened FS pocket is dominated by axial orbital in NNO, the multiband picture is less prominent in LNO. This is also reflected in the far weaker contribution of the inter-orbital susceptibility to be discussed in the following.

The orbital resolved spin susceptibility for NNO is shown in Fig.~\ref{fig:FS}(c), which highlights the importance of inter-orbital
contribution. The relative contributions from axial orbital ($s$) and inter-orbital ($s$-$d$), compared to Ni-$d$ are found to be 1/10-th
and 1/5-th, respectively. In comparison, in LNO, they are 1/100-th and 1/20-th, respectively. This makes the total susceptibility dominated
almost entirely by the $d$-orbital contribution for LNO, while the significant inter-orbital orbital contribution makes the total susceptibility in NNO
appreciably different from the $d$-orbital contribution (cf. Fig.~\ref{fig:FS}(c) and (f)).


\begin{figure}[t]
\includegraphics[width=0.5\textwidth]{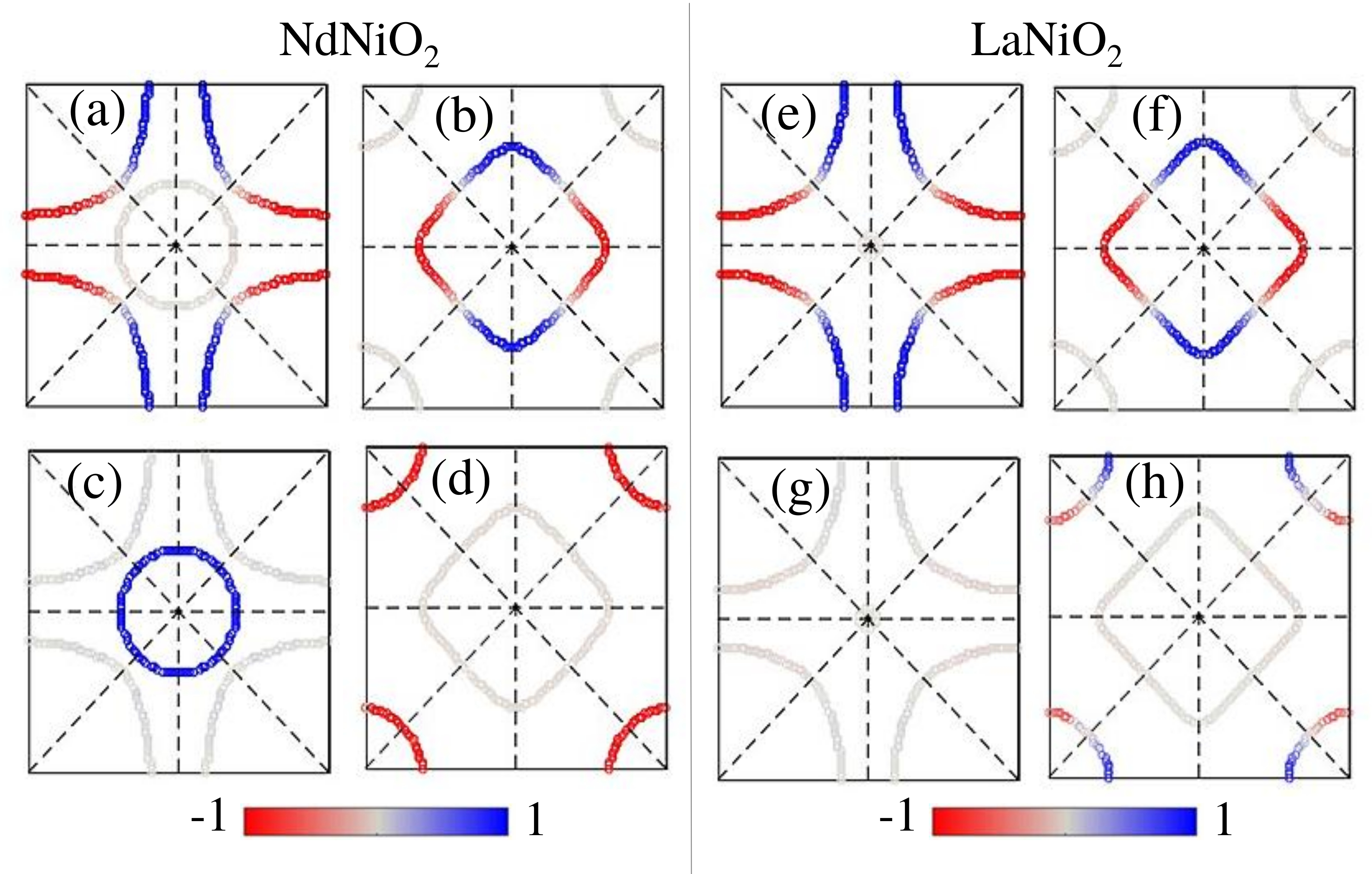}
\caption{(Color online) Computed values of orbital resolved pairing eigenfunction $\Delta_{\alpha}({\bf k})$ plotted on the FS at two representative cuts $k_z= 0$ [(a)-(c)], $\pi$[(b)-(d)] for NNO. [(a),(b)] and [(c),(d)] give
  orbital contributions for Ni $x^{2}$-$y^{2}$ and axial orbital, respectively. (e)-(h) Same as (a)-(d), but plotted for LNO.}
\label{fig:symmetry} 
\end{figure}

In Fig.~\ref{fig:symmetry} we plot the pairing eigenfunction $\Delta({\bf k})$ for the highest eigenvalue $\lambda$, but projected onto the different orbital channels as $\Delta_{\alpha\beta}=\sum_{\nu}\Delta_{\nu}\phi_{\nu}^{\alpha *}\phi_{\nu}^{\beta}$ (${\bf k}$ dependence is suppressed for simplicity), where $\alpha$, $\beta$ are orbital indices, and $\nu$ is the band index. $\phi_{\nu}^{\alpha}$ is the eigenvector of the two-band Hamiltonian. In NNO, we clearly observe that the pairing symmetry of the Ni
$d$ orbital onto the FS is a pure $d_{x^2-y^2}=\cos{k_x}-\cos{k_y}$ type, with very little or no three dimensional component. On the other hand, the projected pairing symmetry on the axial orbital
can be described by a simple $k_z$ dispersion as $\cos{k_z}$, with no signature of the basal plane anisotropy.
In contrast, in LNO compound, the axial orbital's contribution on the FS is drastically reduced, and hence the calculated pairing symmetry changes to a simple $d_{x^2-y^2}$. This result implies that the axial orbital, although seemingly has reduced weight on the FS, can play important role to determine the pairing symmetry. 

\begin{figure}[t]
\includegraphics[width=0.3\textwidth]{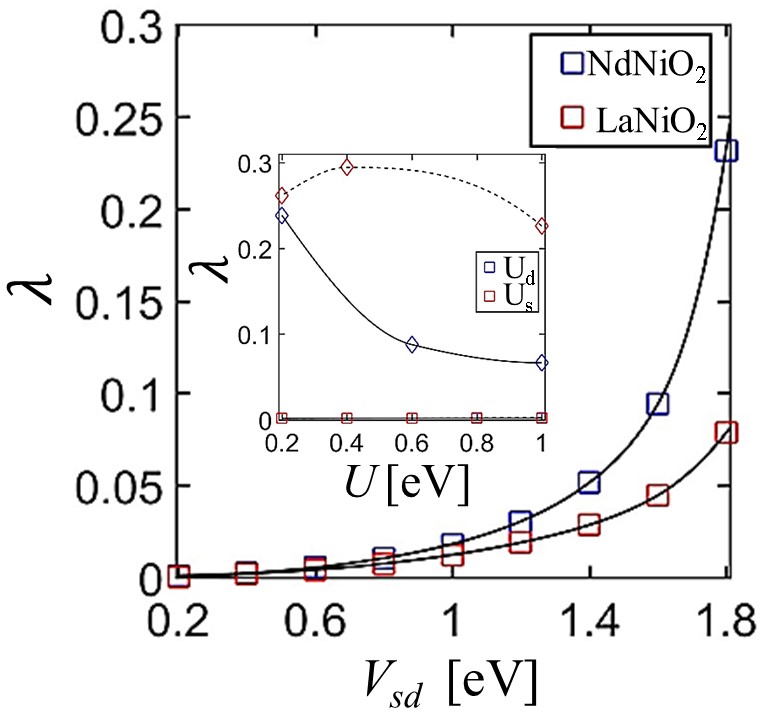}
\caption{Evolution of SC coupling constant $\lambda$ as a function of inter-orbital Hubbard interaction $V_{sd}$ for choice of $U_d$ = 0.9 eV and 0.6 eV for NNO and LNO. Inset shows the variation of $\lambda$ as a function of intra-orbital interactions $U_s$ (red), $U_d$ (blue) for NNO for choice of $V_{sd}$ = 0.5 eV (square) and $V_{sd}$ = 1.5 eV (diamond).} 
\label{fig:lambda} 
\end{figure}

Finally, we study how the pairing strength $\lambda$ depends on the choice of Hubbard interaction parameters, $U_s$, $U_d$
and $V_{sd}$, which unravels as interesting scenario. Firstly, focusing on NNO,  we find that $\lambda$ increases almost exponentially with $V_{sd}$ (cf. Fig.~\ref{fig:lambda}), while neither $U_d$ or $U_s$ is effective in enhancing $\lambda$.  Thus, an appreciable
$\lambda$ is obtained only when $V_{sd}$ becomes appreciable. Secondly, relative to NNO, the pairing strength grows much more
slowly with$V_{sd}$  in LNO. Thus  even for appreciable value of $V_{sd}$, the
the pairing strength in LNO is much smaller than NNO. This in turn highlights the important role of the inter-orbital
interaction $V_{sd}$ for superconductivity in nickelate compounds under discussion, and their material dependence. 

{\it Conclusion.--} In summary, motivated by the two band scenario,\cite{sawatzky,Vishwanath,hwang-new} proposed for $R$NiO$_2$ ($R$ = La, Nd), we derived a two band Hamiltonian out of DFT calculations, keeping the Ni $x^{2}$-$y^{2}$ and Ni $s$ degrees of freedom active, and
integrating the rest. The latter forms an axial orbital from a combination of Nd/La $d$, Ni 3$z^{2}$-$r^{2}$ and Ni $s$, and encodes
the materials dependence. Calculation of superconducting properties in such a two orbital picture, shows an orbital selective pairing
for the Nd compound, while it is found to be only of $x^2$-$y^2$ symmetry in La compound. Most importantly, we find while the SC pairing
grows almost in an exponential fashion with inter-orbital Hubbard interaction for the Nd compound, it is not helped by the choice of
intra-orbital Hubbard interactions. Though the same holds good for La compound, the growth of pair interaction with the strength of
inter-orbital Hubbard interaction is much weaker than in Nd compound, presumably justifying the fact that superconductivity has
been so far observed only for the Nd compound.\cite{hwang, hwang-new} 

\acknowledgements
T.S-D acknowledges financial support from Department of Science and Technology, India. T.S-D and I.D acknowledge support from the National Science Foundation under Grant No. NSF PHY-1748958 and hospitality from KITP where part of this work was performed. TD acknowledges supports from the MHRD, Govt. of India under STARS research funding. 

$\ast$ Equal contribution.


\begin{thebibliography}{99}
\bibitem{footnote1}Since the weigth of the axial orbital largely belongs to La/Nd $d$, the Hunds coupling is not included.
\end{thebibliography}
\end{document}